\documentclass[%
 reprint,
 amsmath,amssymb,
 aps,
]{revtex4-1}

\usepackage{upgreek}
\usepackage{graphicx}
\usepackage{dcolumn}
\usepackage{bm}
\usepackage{hyperref}

\newcommand{\ppcLink}[2]{\hyperref[#2]{#1~\ref*{#2}}}
\newcommand{\ppcBraKet}[2]{\ensuremath{\left\langle{#1}\mathrel{\left|{\vphantom {#1 #2}}\right.\kern-\nulldelimiterspace}{#2} \right\rangle}}
\newcommand{\ppcKet}[1]{\ensuremath{\left|{#1}\right\rangle}}

\newcommand{\ppcFigureScaleA}[4]{\begin{figure}[!t]\begin{center}\includegraphics[width=#4\columnwidth]{#1}\end{center}\caption{#2}#3\end{figure}}

\newcommand{\unit}[1]{\,\mathrm{#1}}

\newcommand{\figLabel}{Fig.}

\begin{document}


\title{Single-shot single-gate RF spin readout in silicon}


\author{P. Pakkiam}
\author{A. V. Timofeev}
\author{M.G. House}
\author{M.R. Hogg}
\author{T. Kobayashi}
\author{M. Koch}
\author{S. Rogge}
\author{M.Y. Simmons}%
\affiliation{%
 CQC2T, UNSW, Sydney, Australia
}%



\date{\today}

\begin{abstract}
For solid-state spin qubits, single-gate RF readout can help minimise the number of gates required for scale-up to many qubits since the readout sensor can integrate into the existing gates required to manipulate the qubits~\cite{Veldhorst_2017,Pakkiam_2018}. However, a key requirement for a scalable quantum computer is that we must be capable of resolving the qubit state within single-shot, that is, a single measurement~\cite{DiVincenzoCriteria}. Here we demonstrate single-gate, single-shot readout of a singlet-triplet spin state in silicon, with an average readout fidelity of $82.9\%$ at a $3.3\unit{kHz}$ measurement bandwidth. We use this technique to measure a triplet $T_-$ to singlet $S_0$ relaxation time of $0.62\unit{ms}$ in precision donor quantum dots in silicon. We also show that the use of RF readout does not impact the maximum readout time at zero detuning limited by the $S_0$ to $T_-$ decay, which remained at approximately $2\unit{ms}$. This establishes single-gate sensing as a viable readout method for spin qubits.
\end{abstract}

\pacs{Valid PACS appear here }
\maketitle


\section{\label{sec:intro}Introduction}

Semiconductor quantum dots show great potential for scalable quantum information processors~\cite{Loss_1998,Levy_2002,Hanson_2007,Zwan_2013}. Singlet-triplet qubits, formed by taking the subspace of the two-electron spin states singlet $S_0=(\ppcKet{\uparrow\downarrow}-\ppcKet{\downarrow\uparrow})/\sqrt{2}$ and triplet $T_0=(\ppcKet{\uparrow\downarrow}+\ppcKet{\downarrow\uparrow})/\sqrt{2}$ under a energy gradient (such as a magnetic field gradient across two quantum dots), have enabled all electrical control of qubit rotations while demonstrating immunity to common mode magnetic field noise~\cite{Petta_2005,Shulman_2012}. The singlet-triplet subspace spanned by $S_0$ and triplet $T_-=\ppcKet{\downarrow\downarrow}$ can also be used to read out single electron spins. Here by loading a spin-down electron onto the dot with the lower spin-down ground state, RF readout can be used to measure the spin state of the target electron on the other dot~\cite{Veldhorst_2017}. If the target electron on the other dot is spin-down, Pauli blockade prevents the target electron from tunnelling across the dots and yields no RF response, while a spin-up electron will form a singlet state with the other electron, giving a non-zero RF response. One of the challenges in scaling up to many qubits is the space real-estate needed for the spin sensors required to readout and initialise the individual qubits. An optimal solution has been suggested to use the mandatory gates assigned for qubit control and manipulation as single-gate RF sensors~\cite{doi:10.1021/nl100663w,Colless_2013,House_2015,Veldhorst_2017,Pakkiam_2018}. To date, however the sensitivity of such single gate sensors has not been high enough to achieve single-shot readout. Single-shot qubit readout is a necessary requirement for running error correction codes where time-correlated measurements are required between many qubits~\cite{DiVincenzoCriteria,PhysRevA.86.032324}. 

Phosphorus donor quantum dots have previously exhibited large ($\sim8\unit{meV}$) singlet-triplet splittings~\cite{Weber_2014}, with independent readout of double quantum dot systems using three-lead single electron transistor (SET) sensors~\cite{Broome2018,Watsone1602811}. When replacing the three lead sensors with a single gate sensor, large $S_0$ to $T_-=\ppcKet{\downarrow\downarrow}$ relaxation times of $2\unit{ms}$ have been achieved, enabling sufficient integration times to perform spin state readout~\cite{Pakkiam_2018}. However, the sensitivity of the resonator circuit was limited by the low quality factor of its \emph{Coilcraft 1206CS-821XJE} chip inductor. Superconducting inductors have recently demonstrated effective quality factors of up to 800, subsequently increasing the sensitivity of the readout circuit~\cite{Colless_2012,PhysRevApplied.9.054016,PhysRevApplied.10.014018}.

\section{\label{sec:method}Method}

In this paper we integrate a superconducting inductor into a single gate donor-based quantum dot architecture in silicon for single-shot readout. The device shown in \ppcLink{\figLabel}{fig:device}a (previously measured in~\cite{Pakkiam_2018}) was fabricated in silicon with the leads and dots defined by atomically placed phosphorus donors using hydrogen resist scanning tunnelling microscope (STM) lithography~\cite{Fuechsle_2012}.
Two pairs of quantum dots (D1L, D1U) and (D2L, D2U), each consisting of approximately 3-4 donors each, are each manipulated by two leads: a reservoir to load electrons and a gate to tune the singlet-triplet state. Single-shot readout was performed on a singlet-triplet state hosted across the dots D2L and D2U, using the resonator connected to reservoir R2. A global tunnel junction charge sensor TJ was patterned at the side and connected to a chip inductor resonator to help locate a singlet-triplet charge transition. The resonators were connected to a frequency multiplexed line~\cite{Hornibrook_2014,House_2016,Pakkiam_2018}.

We have incorporated a $100\unit{nm}$ thick NbTiN, on Si subtrate, superconducting spiral inductor on the single-gate sensor R2 to increase the quality factor for maximal readout signal (as shown in \ppcLink{\figLabel}{fig:device}b). This inductor was a 14-turn spiral, $78\unit{mm}$ in length, $10\unit{\upmu m}$ in width and had a $30\unit{\upmu m}$ gap between turns. The inductor was found to retain its effective quality factor in parallel magnetic fields of $\sim3.3\unit{T}$. These large fields are necessary for both operating singlet-triplet qubits (to break the triplet degeneracy) and in performing RF readout where the $S_0$-$T_-$ energy anti-crossing should not intersect the RF tone as this would cause the qubit state to change during the measurement due to singlet triplet $T_-$ mixing~\cite{Gorman2018}. \ppcLink{\figLabel}{fig:device}c shows the frequency response of the inductor when connected to R2. The internal and external quality factors of this inductor when wire-bonded to the device ($75\unit{mK}$), were approximately 800 and 400 respectively. The resonator's frequency was $339.6\unit{MHz}$ at zero magnetic field and $335.2\unit{MHz}$ at $2.75\unit{T}$.

\ppcFigureScaleA{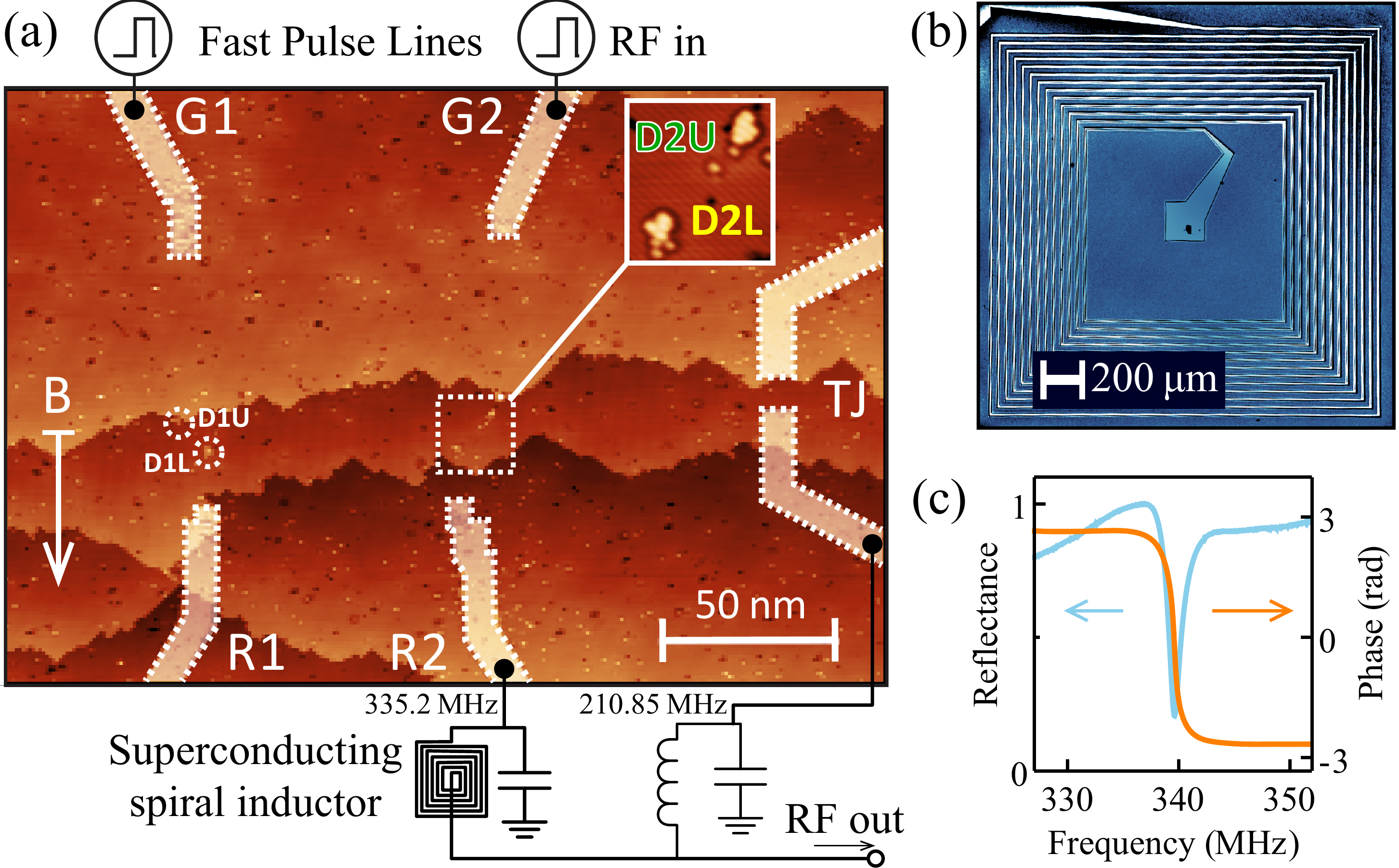}{\textbf{Single gate sensor readout with integrated superconducting resonator circuit}. \textbf{(a)}~The STM image shows the silicon surface lithography where the lighter regions have been desorbed from the lithographic hydrogen mask. These areas are dosed with phosphorus to form the dots (D1U, D1L, D2U and D2L) and metallic electrodes~\cite{Weber_2012}. A standard chip-inductor resonator connects to the tunnel junction charge sensor TJ. Reservoir R2 is used to load dots D2L and D2U with electrons while the gates G1 and G2 are used to manipulate the singlet-triplet detuning of the dot pairs. B indicates the in-plane magnetic field during millikelvin measurements. \textbf{(b)}~The superconducting resonator is added to the frequency multiplexed line, connected to R2 and measures the singlet-triplet state across D2U and D2L. \textbf{(c)}~The reflected (sending and receiving the RF tone via the multiplexed line) frequency response of the superconducting inductor when connected to R2 at zero magnetic field.}{\label{fig:device}}{0.95}

\section{Results}

In \ppcLink{\figLabel}{fig:pwrOpt}a we plot the differential response from the tunnel junction charge sensor TJ as we sweep the gates G1 and G2 at the $(3,3)$ to $(2,4)$ inter-dot charge crossing. This charge configuration is equivalent spin-wise to a $(1,1)$-$(0,2)$ singlet-triplet crossing where we observe a clear inter-dot transition due to the tunnelling of a single electron. We measure the tunnel-coupling at the $(3,3)$ to $(2,4)$ transition as $39\pm6\unit{GHz}$ by plotting the dependence of the inter-dot transition at different applied magnetic fields~\cite{Pakkiam_2018}. Since the tunnel coupling is much larger than the driving frequency of the resonator ($335.2\unit{MHz}$), this inter-dot transition forms a good candidate for single-gate readout as the RF drive will ensure that the electron will adiabatically oscillate between the two dots.

\ppcFigureScaleA{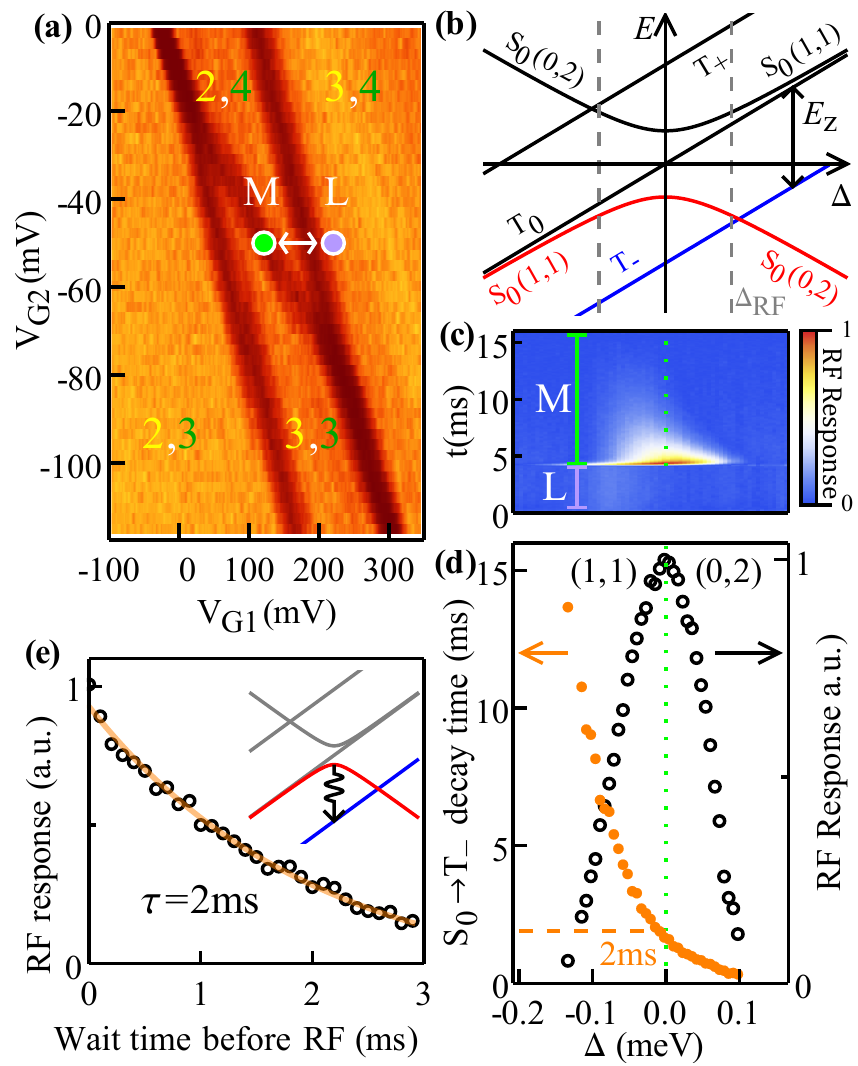}{\textbf{Optimising the singlet-triplet readout position}. Using the sensor on R2, singlet states were time-resolved to find the optimal point in detuning. \textbf{(a)}~Charge stability diagram around the $(2,4)$-$(3,3)$ transition taken by observing the differential response of the sensor TJ whilst sweeping the gates G1 and G2 across a singlet-triplet transition on D2L and D2U. \textbf{(b)}~The singlet-triplet energy diagram highlighting the $T_-$ ground state (blue) and the $S_0(1,1)$ to $S_0(0,2)$ crossing (red) where the electron oscillates when performing singlet readout. $E_\text{z}$ is the Zeeman of the triplet $T_\pm$ states due to the applied $2.75\unit{T}$ magnetic field, while $\Delta_\text{RF}$ denotes the amplitude of the RF tone. \textbf{(c)}~RF response averaged over 10,000 shots at different points in detuning. Each trace was taken when waiting at L for $4.1\unit{ms}$. The non-zero RF response signifies the presence of oscillating electrons (singlet states) that eventually decay into triplet $T_-$ states. \textbf{(d)}~When fitting an exponential to each trace, it is clear that at zero detuning, the RF response is maximal. \textbf{(e)}~In all the experiments (a-d), the RF tone was constantly present. In this plot, we show the resulting RF response (fitted from the exponential $S_0$-$T_-$ decays as in c) at zero detuning when turning on the RF pulse after a time waiting $0$-$3\unit{ms}$ at M. The resulting decay (shown in the inset) has the same time constant of $2\unit{ms}$ suggesting that the decay is not an effect of applying the RF pulse.}{\label{fig:pwrOpt}}{0.9}

When performing readout, the electrons oscillate between the dots giving rise to a measurable quantum capacitance~\cite{doi:10.1021/nl100663w,Colless_2013,House_2015,Pakkiam_2018}. Triplet states cannot oscillate electrons due to Pauli blockade. Singlet states adiabatically move one electron between the dots, the $S_0(1,1)$ and $S_0(0,2)$ states, shown by the red branch in \ppcLink{\figLabel}{fig:pwrOpt}b. The optimal point for maximal electron shuttling is at zero detuning $\Delta=0$, since here the RF tone moves the electron the greatest distance, pushing it equally into both dots respectively. We set the magnetic field to $2.75\unit{T}$ as this moves the $S_0$-$T_-$ anti-crossing (the overlap of the red and blue lines) far enough away from the zero-detuning point such that the RF tone does not intercept this anti-crossing. We test the response of the single gate sensor on R2 using the multi-purpose fast-pulse gate, G2. Here we send the RF tone with an amplitude $\Delta_\text{RF}$ (grey dotted lines) through G2 while the superconducting resonator on R2 captures the resulting RF response. The response was fed into a lock-in amplifier, which modulated the amplitude of the input RF tone at $21.361\unit{kHz}$, to filter the detection of noise originating from the room temperature apparatus. The overall measurement bandwidth was approximately $3.3\unit{kHz}$.

If we consider \ppcLink{\figLabel}{fig:pwrOpt}a, we can load a singlet $S_0(0,2)$ state by pulsing from the $(3,4)$ state at point L to the $(2,4)$ state at point M. It is important that we wait $\sim4\unit{ms}$ at point L before moving to the measurement point M so that the spin relaxes to the singlet ground state as discussed later. The energy diagram describing the two electrons across the two quantum dots (\ppcLink{\figLabel}{fig:pwrOpt}b) shows that at the zero detuning readout point M, the triplet $T_-$ (blue line) is the ground state. Thus, the singlet $S_0$ will eventually decay into the triplet $T_-$ state (inset \ppcLink{Fig.}{fig:pwrOpt}e) during measurement and this sets an upper bound to the overall measurement time.

To find the optimal detuning point for maximal response, we measure the RF response at different points in detuning, as shown in \ppcLink{\figLabel}{fig:pwrOpt}c, taking an average of 10,000 individual time traces at each point. On moving point M from negative detuning to zero detuning, a non-zero response is observed indicating the presence of a singlet state. This signal decays again at positive detuning where the singlet population decreases as the RF tone oscillates past $T_-$-$S_0(0,2)$ anti-crossing~\cite{Gorman2018}. We fit these decay events at different points in detuning $\Delta$ to an exponential distribution giving the amplitudes and time constants of the decay events in \ppcLink{\figLabel}{fig:pwrOpt}d (in black and orange respectively). It is clear that the optimal RF response occurs at zero detuning as the electrons are maximally shuttled between $S_0(1,1)$ and $S_0(0,2)$ states.

Finally we test whether the use of the RF tone itself shortens the $S_0$-$T_-$ lifetime (for example, due to spin-orbit coupling). This would be undesirable since we do not want the detector to affect the population dynamics of the singlet-triplet state during measurement. To measure the bare $S_0$-$T_-$ lifetime when no RF tone is applied we started with the RF tone turned off, waited at point L over $4.1\unit{ms}$ (to load a singlet as before) and then moved to zero detuning. We only switched on the RF tone to measure the singlet population after waiting different time-periods at point M, as shown in \ppcLink{\figLabel}{fig:pwrOpt}e. When fitting to the resulting exponential decay of the singlet population, we found that the decay time remained the same ($2\unit{ms}$) as that when the RF tone remained switched on during the whole experiment (\ppcLink{\figLabel}{fig:pwrOpt}d). Thus, we conclude that the RF excitation does not play a major role in the $S_0$-$T_-$ decay as the singlet lifetime remains unaffected by the RF measurement tone.

\ppcFigureScaleA{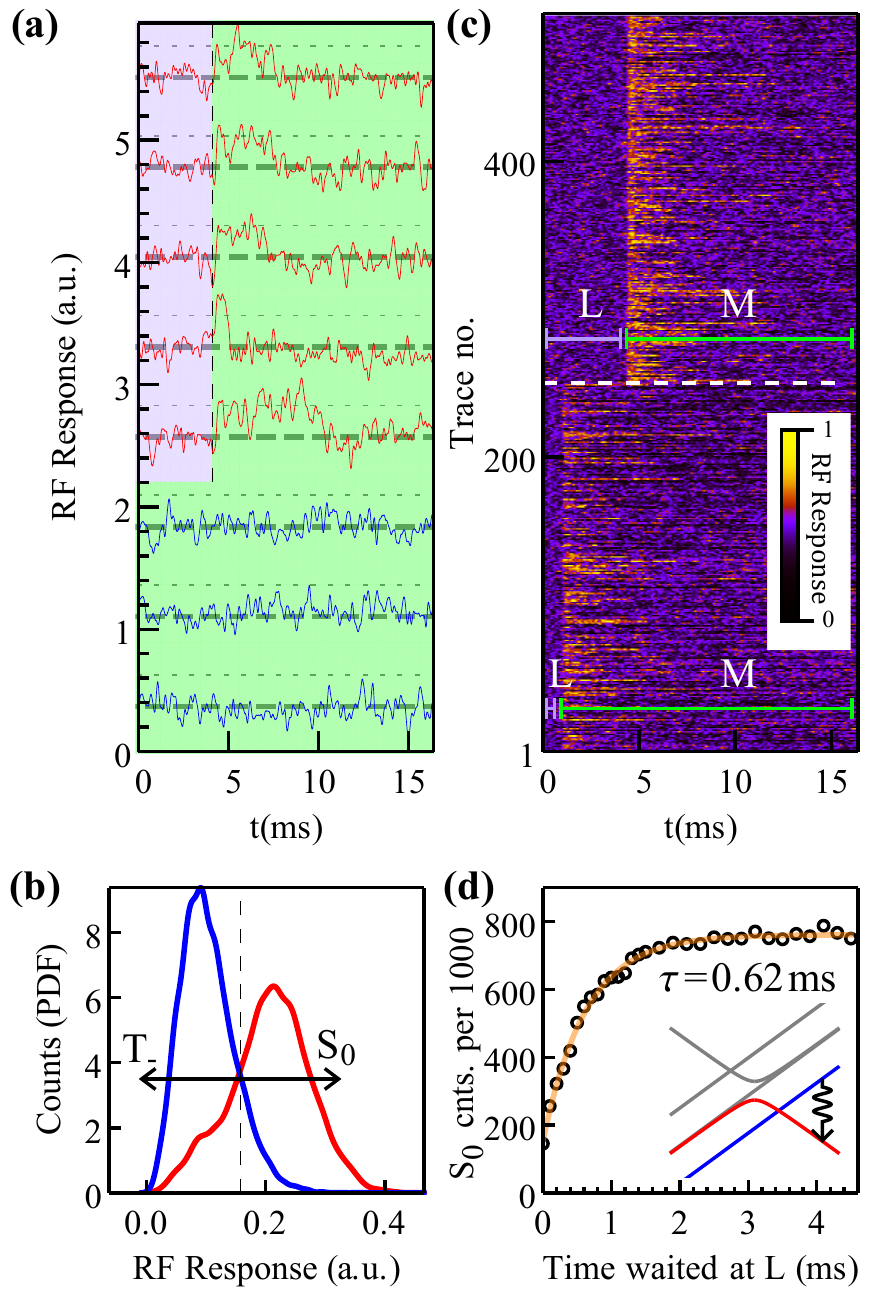}{\textbf{Single-shot single-spin readout}. Using the superconducting spiral inductor on our single gate sensor R2, single spin states could be resolved with single shot. \textbf{(a)}~Single-shot traces (offset) measuring a single spin in the singlet state when waiting at L for $4.1\unit{ms}$ (shown in red) compared to those in triplet states where we did not pulse to L (shown in blue). \textbf{(b)}~A histogram (a probability density function (PDF) from 10,000 traces) of the maximum value of the RF response when waiting at point L for $0\unit{s}$ and $4.1\unit{ms}$ shown in blue and red respectively. The dashed line shows the selected threshold that maximises the readout fidelity at $82.9\%$. \textbf{(c)}~500 individual time traces of the RF response. The first 250 were taken after waiting at L for $0.7\unit{ms}$ to partially load singlet states and the second 250 traces after waiting at L for $4.1\unit{ms}$ to fully load singlet states. The high RF response signifies the presence of singlet states that stochastically decay into triplet $T_-$ states. The shorter wait time highlights the lower singlet population as insufficient time was given for the $T_-$ state to decay into the $S_0$ state before measurement. \textbf{(d)}~To observe this dependence, using the optimal readout threshold, 1000 single-shot traces were taken to measure the singlet population on varying the time spent at point L. This probes the $T_-$ to $S_0$ relaxation time at $\Delta\sim1\unit{meV}$ of $0.62\unit{ms}$.}{\label{fig:ssTraces}}{0.9}

We perform single-shot measurements of spin-up electrons (equivalent to the singlet state) by waiting at point L for $4.1\unit{ms}$ and then pulsing to point M at zero detuning for maximum RF response. Here we leave the RF tone switched on and measure the resulting response over time in \ppcLink{\figLabel}{fig:ssTraces}a. Five such single-shot time-traces are shown in red demonstrating that we can detect a singlet state. Here, when moving to point M, the signal (dotted lines) clearly departs from the background level (dashed lines) and, after stochastic relaxation, returns back to the same level highlighting real-time single spin detection. The finite life-time of the non-zero signal can be attributed to the singlet-state $S_0$ decaying into the triplet $T_-$ ground state. To distinguish a singlet state from a triplet state, these traces were compared against traces (shown in blue) when waiting $0\unit{s}$ at point L (that is, always reading a triplet $T_-$) where the signal remains at the background level throughout the measurement. To quantify the fidelity of the single spin readout, we must discriminate between a fully null signal (triplet) and one with a non-zero signal (singlet). The singlet signal on average follows an exponential decay function (as shown in \ppcLink{\figLabel}{fig:pwrOpt}e). Therefore, the actual signal is concentrated around the beginning of the measurement and diminishes during the course of the measurement. Thus, we applied an exponential window over the portion of the signal where the measurement begins (after we have moved to zero detuning at point M) and then compiled a histogram of the maximum values of each trace~\cite{PhysRevA.89.012313}. The histogram shown in \ppcLink{\figLabel}{fig:ssTraces}b was created from 10,000 traces taken after waiting at point L for $4.1\unit{ms}$  and $0\unit{s}$ to measure the distribution for singlets and triplets respectively. We took a threshold (shown by the dotted line in \ppcLink{\figLabel}{fig:ssTraces}b) to partition the distributions such that values above were considered to be singlet states and values below were considered to be triplet states. We determined that the optimal threshold to maximise the readout fidelity was 0.157~\cite{Elzerman_2004,PhysRevApplied.8.034019}. This yielded an average single-spin readout fidelity, of the single-gate RF sensor, of $82.9\%$ (where the singlet and triplet readout fidelities were $78.2\%$ and $87.6\%$ respectively).

With the ability to perform single-shot, single-spin readout, we measure the $T_-$ to $S_0(0,2)$ decay (at $\sim1\unit{meV}$ from zero detuning) by varying the time the pulse spends at point L. After every measurement, the electrons will have decayed into the triplet $T_-$ state. On initially pulsing to point L, the electrons (residing in separate dots) remain in $T_-$ instead of entering the $(3,4)$ charge state. This is because there is a slower tunnel rate of electrons from the reservoir R2 to D2U since it is further away when compared from R2 to D2L. Thus, the electrons cannot move to the $(3,4)$ charge state until the electron in D2L moves to D2U to enter the singlet $S_0$ state~\cite{yang_2014,Pakkiam_2018}. \ppcLink{\figLabel}{fig:ssTraces}c shows 250 traces taken when waiting $0.7\unit{ms}$ at point L and 250 traces when waiting $4.1\unit{ms}$ at point L. The RF response starts to appear when measuring at point M. The lengths of each non-zero signal are exponentially distributed with a time constant of $2\unit{ms}$ and represent singlet states decaying into triplet $T_-$ states. When waiting a lower time at L, there is clearly a smaller proportion of singlet states. \ppcLink{\figLabel}{fig:ssTraces}d shows the singlet-counts over 1000 traces taken at different wait times at point L. When viewing the singlet-counts as a function of the wait time at point L and fitting to the resulting exponential rise in the singlet counts, the decay time was measured to be $0.62\unit{ms}$. This corresponds to the upper bound in the measurement time when using a charge sensor to distinguish between $T_-(1,1)$ and $S_0(0,2)$ states~\cite{Shulman_2012}.


\section{\label{sec:conclusion}Conclusion}

In summary we demonstrated single shot readout of an electron spin using the singlet-triplet basis in silicon using a single-gate RF sensor. The reduction in gate density using a single-gate sensor simplifies architectures for large arrays of solid state qubits~\cite{Veldhorst_2017,Pakkiam_2018}. We demonstrated that the $S_0$ to $T_-$ relaxation time, which limits the qubit measurement time, was $2\unit{ms}$ and unaffected by the presence of the RF tone. The single-gate RF sensor gave an average measurement fidelity of $82.3\%$. The fidelity can be further improved by using resonators with higher internal quality factors along with better matching networks to the transmission line for greater signal strength. For example, in this experiment, if the internal and external quality factors were matched at 1600 the expected readout fidelity exceeds $99\%$.

\section*{Acknowledgements}

This research is supported by the Australian Research Council Centre of Excellence for Quantum Computation and Communication Technology (Project No. CE110001027) and the U.S. Army Research Office under Contract No. W911NF-17-1-0202. The device is fabricated in part at the New South Wales node of the Australian National Fabrication Facility. M.Y.S. acknowledges an Australian Research Council Laureate Fellowship.

\bibliography{references}

\end{document}